\RequirePackage{fix-cm}
\documentclass[smallextended]{svjour3}       
\smartqed  
\usepackage{graphicx}

\newtheorem{prop}{Proposition}

\usepackage{algorithm}

\usepackage{subfigure,float}
\usepackage{booktabs}
\usepackage{amsmath,amssymb}
\usepackage{color,soul}
\usepackage{enumerate}
\usepackage{hyperref}
\usepackage{latexsym}
\usepackage{graphicx,graphics}
\usepackage{tabu}
\usepackage{longtable}
\usepackage{natbib}
\DeclareGraphicsExtensions{.bmp,.jpg,.eps,.pdf,.png}
\usepackage{float,verbatim}

 \journalname{....}
\begin{document}

\title{Isomorphism Check for $2^n$ Factorial Designs with Randomization Restrictions
	\thanks{Ranjan's research was partially supported by the IIM Indore's Grant for External Research Collaboration. Mendivil's research was funded in part by NSERC 2012:238549.}
}


\author{Neil A. Spencer         \and
        Pritam Ranjan  \and Franklin Mendivil
}

\authorrunning{Spencer, Ranjan and Mendivil} 

\institute{Neil A. Spencer \at
Department of Statistics and Data Science\\
              Carnegie Mellon University, Pittsburgh, USA \\
              \email{nspencer@andrew.cmu.edu}           
           \and
           Pritam Ranjan (corresponding author) \at
              Operations Management \& Quantitative Techniques Area\\ Indian Institute of Management Indore, India\\
              Phone: +91 7312439512\\
              \email{pritamr@iimidr.ac.in} 
                         \and
           Franklin Mendivil \at
              Department of Mathematics and Statistics\\
               Acadia University, Canada\\
              \email{franklin.mendivil@acadiau.ca}   
}

\date{Received: date / Accepted: date}

\maketitle

\begin{abstract}

Factorial designs with randomization restrictions are often used in industrial experiments when a complete randomization of trials is impractical. In the statistics literature, the analysis, construction and isomorphism of factorial designs has been extensively investigated. Much of the work has been on a case-by-case basis -- addressing completely randomized designs, randomized block designs, split-plot designs, etc. separately. In this paper we take a more unified approach, developing theoretical results and an efficient relabeling strategy to both construct and check the isomorphism of multi-stage factorial designs with randomization restrictions. The examples presented in this paper particularly focus on split-lot designs.


\end{abstract}

\section{Introduction}
\label{sec:intro}

Factorial designs are common in a wide variety of applications, however, complete randomization of trials is often impractical. This could be because some factors may be more expensive to change than others, the trials may need to be partitioned into homogeneous batches at each stage, the experimental units may need to be processed multiple times under different settings, and so on. Popular factorial designs with randomization restrictions include blocked designs, split-plot designs, strip-plot designs, split-lot designs, and combinations thereof. Though both the construction and the analysis of such designs have been active areas of research for decades, most of the literature focuses on a case-by-case basis. The literature started with the exploration of completely randomized designs (CRDs) and randomized block designs (RBDs) by R. A. Fisher and F. Yates. Later, \cite{addelman1964some}, \cite{bingham1999minimum} and many others investigated split-plot designs, \cite{miller1997strip} pioneered strip-plot designs, and \cite{mee1998split} and \cite{butler2004} presented some fundamental results on split-lot designs. See \cite{dean1999design}, \cite{mukerjee_wu_book}, \cite{hamada1992analysis},  \cite{hedayat2012orthogonal}, and \cite{cheng2016theory} for detailed references. In this paper, we focus on multi-stage factorial designs with randomization restrictions under a unified framework.

For easier understanding and more concise notation, we concentrate on two-level factorial designs, however, several results and algorithms presented in this paper can easily be extended for $q$-levels. Consider a factorial experiment investigating the significance of $n$ basic factors and all of their interactions. Each $r$-factor interaction can be expressed as an $n$-dimensional vector composed of exactly $r$ ones and $n-r$ zeros, with the $r$ ones indicating which basic factors are present in the interaction term. We employ the following shorthand. Let the first $n$ uppercase letters $A, B, C,\ldots$ denote the $n$ basic factors in the experiment, and denote any interaction as a string of letters composed of the basic factors that are involved. For instance, $\{A,B,AB,C,...,ABCDE\}$ denotes all basic factors and their interactions for a $2^5$ factorial experiment, with $AE = (1,0,0,0,1)$ representing a two-factor interaction between the first and the fifth basic factors.

Multi-stage factorial experiments are common in industrial applications, where all experimental units are processed at each stage, and the observations are taken at the end after the final stage. Some traditional designs like a RBD and a split-plot design can be thought of as multi-stage factorial designs with only one stage. A non-trivial example is a split-lot design (also referred to as \emph{multiway split-unit design}), which consists of multiple processing stages with each stage using a split-plot design to partition the experimental units \citep{ryan2007modern}. Popular applications include the laundry experiment for measuring wrinkles in \cite{miller1997strip} and \cite{mee2009comprehensive}, the fabrication of integrated circuits using silicon wafers in \cite{mee1998split}, and the plutonium alloy experiment in \cite{bingham2008factorial}. See Section~2.1 for more details on some of these examples.

Over the decades, a handful of unifying methodologies have been developed for studying different factorial designs with randomization restrictions. For instance, \cite{nelder_1965a, nelder_1965b} developed the notion of the simple block structure;  \cite{speed_bailey_1982} and \cite{tjur_1984} used association schemes to extend the simple block structure idea to orthogonal block structures which are more powerful and can characterize a wide variety of designs (see \cite{bailey2004association} and \cite{cheng2011paper} for more details). 
In this paper, we use the unified theory proposed by \cite{ranjan2007factorial} which is inspired from the randomization group idea of \cite{bingham2008factorial}. 
The comparison of different unified frameworks is outside the purview of this paper and we do not claim that the the unified theory considered here is more general or powerful.

The key idea behind the unified theory of \cite{ranjan2007factorial} is to realize that the set of all factorial effects (main effects and all possible interactions) of a $2^n$ factorial design constitute an $(n-1)$-dimensional finite projective geometry $\mathcal{P}_n := PG(n-1, 2)$ over $GF(2)$, which is the same as the $n$-dimensional vector space $V(2^n)$ over $GF(2)$  without the zero element \citep{bose1947mathematical}. 
Furthermore, the randomization restrictions for any given stage of a multi-stage factorial design can be characterized by a projective subspace of $\mathcal{P}_n$. 
Such a subspace is referred to as a \emph{randomization defining contrast subspace} (RDCSS). 
Note that a $(t-1)$-dimensional projective subspace (also referred to as a $(t-1)$-flat) of $PG(n-1,2)$ is the same as a $t$-dimensional vector subspace of $V(2^n)$ excluding the zero element. In this approach, the overlapping pattern of the flats were exploited to construct useful split-lot designs. \cite{ranjan2009existence} formalized the construction and analysis of split-lot designs that are derived from the set of disjoint flats of $\mathcal{P}_n$.  
In several real-life application, for instance, in the plutonium alloy experiment of \cite{bingham2008factorial}, the overlap between two RDCSSs (or flats) cannot be avoided. 
Subsequently, \cite{ranjan2010stars}  proposed a new class of split-lot designs which is based on flats with a common overlap. {Though the following work, \cite{ranjan2007factorial}, \cite{ranjan2009existence} and \cite{ranjan2010stars}, have been frequently cited in this article, this is a standalone paper and, we have presented all necessary results for completeness and easy readability.}

The construction and ranking of designs becomes important whenever there are potentially multiple candidate designs meeting the design requirements of a particular factorial experiment. In the design of experiments literature, one can find a plethora of research articles that focus on innovative techniques for constructing good designs, a variety of design ranking criteria (e.g., maximum resolution and minimum aberration), and methods of sorting through the candidates to find different or \emph{non-isomorphic} designs. However, most of the literature operates on a case-by-case basis (e.g., \cite{bingham1999minimum}, \cite{ma2001isomorphism}, \cite{cheng_tang_2005}, \cite{lin_sitter_2008}).

This paper concentrates on the problem of checking the isomorphism of $2^n$ multi-stage factorial designs under the unified framework characterized by the RDCSS structure. 
As per \cite{ma2001isomorphism}, two fractional factorial designs are said to be isomorphic if one can be obtained from the other by relabeling the factors, reordering the runs, and switching the levels of factors. For a multi-stage factorial experiment, \cite{bingham2008factorial} introduces an update to this definition by adding stage-wise restrictions.

In this paper, we present a formal definition of isomorphism using the RDCSS-based unified framework. If performed naively, checking for the isomorphism of designs can involve iterating over all possible relabelings and reorderings, which quickly becomes computationally infeasible for large designs. We present a bitstring representation of $\mathcal{P}_n$ which helps in developing an efficient search algorithm. A new search strategy is also proposed which exploits the geometric structure of RDCSSs to significantly reduce the search space. We further apply known results from projective geometry (e.g. \cite{soicher2000computation, topalova20102, mateva2009line}) to completely classify the isomorphism properties for several RDCSS-based designs (with small runsizes) that are useful from a practical standpoint. 
We also provide a new result which establishes that all RDCSS-based designs constructed using the cyclic method of \cite{hirschfeld1998projective} are isomorphic. These results are useful for determining when several isomorphism classes of designs must be considered. {Furthermore, all proposed algorithms and important functions have been implemented in} \verb"R" and available on GitHub for easy access.

After reviewing the background theory, important existing results, and motivating examples for the RDCSS-based multi-stage factorial experiments in Section~2, the formal definition of equivalence and isomorphism are presented in Section~3. Sections~4 and 5 presents new theories and algorithms for reducing the search space and efficiently iterating through all possible relabelings for an isomorphism check. 
Section~6.1 reviews the classification of such designs from a practical standpoint, and Section~6.2 presents a theoretical result on the cyclic construction of RDCSSs. Finally, the concluding remarks are summarized in Section~7.


\section{Background Review}

A good design is often expected to facilitate efficient analysis. \cite{daniel1959use} suggested that for unreplicated factorial experiments, the significance of factorial effects can be assessed using half-normal plots with the restriction that the effects appearing on the same plot must have the same error variance. 
{Moreover, each half-normal plot should contain at least six or seven factorial effects for a meaningful inference. 
For an RDCSS-based multi-stage design, the variance of an estimator of a factorial effect is characterized by its presence in different RDCSSs. 
That is, if a multi-stage factorial design is defined by $m$ RDCSSs and all RDCSSs are disjoint, then at most $m+1$ half-normal plots are required for the significance assessment of factorial effects. 
On the other hand, if some of the $m$ RDCSSs overlap then more than $m+1$ separate half-normal plots would have to be used for the identification of significant factorial effects.} 
As a result, it is desirable to construct RDCSSs that are big enough (with at least six or seven effects per half-normal plot) and disjoint. 
Section~2.1 presents a quick recap of two popular examples of multi-stage split-lot designs, and Section~2.2 reviews some relevant results from the finite projective geometry literature, \cite{ranjan2007factorial}, \cite{ranjan2009existence} and \cite{ranjan2010stars}, that are helpful for our discussion on isomorphism.

\subsection{Examples of Multi-stage Experiments}
In this section, we recap the silicon wafer example \citep{mee1998split} and plutonium alloy experiment \citep{bingham2008factorial}).

The fabrication of integrated circuits on silicon wafers goes through a sequence of processing steps. \cite{mee1998split} discussed the construction and analysis of several split-lot designs for analyzing this process. Here, we present two designs for a 64-wafer experiment with nine processing stages and six basic factors. At each stage, all experimental units (i.e., 64 wafers) are processed and then passed on to the next stage. The measurements are taken at the end after the final stage, and the randomization of trials at each processing stage is guided by a set of restrictions defined by carefully chosen factors and factor interactions. The designs are given by $IC_1 = \{\langle A, EF, BCE\rangle$, $\langle B, AF, CDF \rangle$, $\langle C, AB, ADE\rangle$, $\langle D, BC, BEF\rangle$, $\langle E, CD, ACF\rangle$, $\langle F, DE, ABD\rangle$, $\langle BD, BF, ACE\rangle$, $\langle AC, CE, BDF\rangle$, $\langle AD, BE, CF\rangle \}$, and $IC_2 = \{\langle A,BD,CF\rangle$, $\langle B,AF,CE\rangle$, $\langle C,BF,DE\rangle$, $\langle D,AC,BE \rangle$, $\langle E,AB,DF\rangle$, $\langle F, AE, CD\rangle$, $\langle AD,BC,EF\rangle$ $\langle ACE,ADF,BEF\rangle$, $\langle ABC,ADE,CEF\rangle\}$, where $\langle \cdots \rangle$ denotes the span of the vectors/effects within, e.g., $\langle A, EF, BCE\rangle = \{A, EF, AEF, BCE, ABCE, BCF, ABCF\}$. For both $IC_1$ and $IC_2$, the nine RDCSSs are disjoint and each of size seven. 
Thus, the significance of all 63 factorial effects (excluding the null) can easily be assessed by nine half-normal plots. 
The question we address here is whether or not the two designs are isomorphic. 
Of course, the ranking of designs is a different question and we leave it for future research.

The so-called ``plutonium alloy experiment" (in \cite{bingham2008factorial}) took place at Los Alamos National Laboratory (LANL), where the objective was to identify the significant factors and factor combinations involved in the process of manufacturing a plutonium alloy cookie which was to be used further for some classified experiments. 
This cookie-making-process involved five basic factors and had to go through three processing stages: casting and two different types of heat-treatments. 
\cite{bingham2008factorial} considered a $2^5$ factorial split-lot design with 32 runs. 
Letting $f^*_i$ denote the flat in $\mathcal{P}_5$ that characterizes the randomization of trials for the $i$-th processing stage, then as per \cite{bingham2008factorial}, the restrictions are: $A, B \in f^*_1$, $C \in f^*_2$ and $D,E \in f^*_3$. Using a computer search, the authors found it impossible to construct disjoint RDCSSs which could facilitate meaningful half-normal plots. The design suggested at the end was, $PA_1=\{\langle A, B, CDE\rangle$, $\langle C, AD, BE\rangle$, $\langle D, E, ABC\rangle\}$, which required four half-normal plots for the significance assessment of effects in $f^*_1, f^*_2, f^*_3$ (excluding the common $ABCDE$) and $\mathcal{P}_5 \backslash \{f^*_1, f^*_2, f^*_3\}$. 
Later on, \cite{ranjan2010stars} recommended an alternative design $PA_2 = \{\langle A, B, DE, ACD\rangle$, $\langle C, AB, DE, ACD\rangle$, $\langle D, E, AB, ACD\rangle\}$ for this experiment. 
As with the previous example, can we check if the two designs $PA_1$ and $PA_2$ are isomorphic?

\subsection{Projective Geometric Structures} 

The questions of the existence and construction of a pre-specified number of disjoint flats of $ \mathcal{P}_n = PG(n-1,2)$ with given sizes are non-trivial. 
The combinatorics literature contains some results on the existence and construction of spreads and maximal partial-spreads of $\mathcal{P}_n$. 

A spread of $\mathcal{P}_n$ is a set of disjoint flats that includes every element of $\mathcal{P}_n$. 
That is, a spread of $\mathcal{P}_n$ is also a cover of $\mathcal{P}_n$. 
A \emph{balanced $(t-1)$-spread} $\psi$ of $\mathcal{P}_n$ consists of only $(t-1)$-flats of $\mathcal{P}_n$. 
For simplicity, we do not consider the unbalanced spread case in this paper. 
A $(t-1)$-spread $\psi$ of $\mathcal{P}_n$ contains $|\psi| = (2^n-1)/(2^t-1)= \sum_{i=1}^{n/t}2^{(i-1)t}$ distinct $(t-1)$-flats that can be used for constructing RDCSSs for different stages of randomization. 
For instance, in the silicon wafers example, $IC_1$ and $IC_2$ are two distinct $2$-spreads of $\mathcal{P}_6$. A necessary and sufficient condition for the existence of a balanced $(t-1)$-spread of $\mathcal{P}_n$ is that $t$ divides $n$ \citep{andre1954nicht}. For instance, in the silicon wafer example, the  existence of a $(3-1)$-spread of $\mathcal{P}_6$ is ensured as $3$ divides $6$.

If $t\nmid n$ (as in the plutonium example), then either a partial $(t-1)$-spread or a non-overlapping set of RDCSSs have to be used for design construction. A partial spread is simply a set of disjoint flats of $\mathcal{P}_n$. 
{Lemma~\ref{lemma:partialspread-1} discusses the existence of a partial $(t-1)$-spread of $\mathcal{P}_n$.
\begin{lemma}[\cite{eisfeld2000partial}] \label{lemma:partialspread-1}
Let $\mathcal{P}_n$ be a finite projective space $PG(n - 1, 2)$, with $n = kt + r$ for $0 < r < t < n$. Then, there exists a partial $(t - 1)$-spread $\psi$ of $\mathcal{P}_n$ with $|\psi| = 2^r\frac{2^{kt}-1}{2^t-1} - 2^r + 1$.
\end{lemma}
}

Assuming the overlap between two RDCSSs cannot be avoided, \cite{ranjan2010stars} proposed a new geometric structure called a star\footnote{Stars were recently reinvented in a collection of works \citep{shaw2014book, mcdonough2014classification} where they are referred to as book spreads.}  which requires all constituent flats to have a common overlap.

\begin{definition}\label{def:star}
A \emph{balanced star}, denoted by $\Omega=St(n, \mu, t, t_0)$, of $\mathcal{P}_n$ is a set of $\mu$ rays ($(t-1)$-flats) and a nucleus (one $(t_0-1)$-flat) in $\mathcal{P}_n$, such that the intersection of any two of the $\mu$ rays is the nucleus (so, $0\le t_0 < t < n$).
\end{definition}

A star $\Omega$ is said to cover $\mathcal{P}_n$ if the combined set of elements of all rays in $\Omega$ is equal to $\mathcal{P}_n$. For the purposes of this paper, all stars are balanced and covering. Thus a star $\Omega=St(n, \mu, t, t_0)$ provides $\mu = (2^{n-t_0}-1)/(2^{t-t_0}-1)$ overlapping RDCSSs of size $2^t -1$ each. For instance, in the plutonium alloy experiment, $PA_2$ represents a $St(5,3,4,3)$. Lemma~\ref{lemma:starspreadcorr} is taken from \citet[Lemma 3]{ranjan2010stars} which establishes the relationship between a spread and a star.

\begin{lemma} \label{lemma:starspreadcorr}
The existence of a balanced covering star $\Omega = St(n,\mu,t,t_{0})$ of $\mathcal{P}_n=PG(n-1,2)$ is equivalent to the existence of an $(h-1)$-spread $\psi$ of $\mathcal{P}_u$, where $u = n - t_0$, and $h=t-t_0$.
\end{lemma}

The proof of Lemma~\ref{lemma:starspreadcorr} easily follows from the following construction steps. Let $ \{f_1,...,f_{\mu}\}$ be the constituents of an $(h-1)$-spread $\psi$ of $\mathcal{P}_u$. 
Then there exists a $(t_0-1)$-flat $\pi$ (referred to as the nucleus) in $\mathcal{P}_n \backslash \mathcal{P}_u$ such that $f_i^* = \langle f_i, \pi\rangle$ and $\{f_1^*,...,f_{\mu}^*\}$ form a covering star $\Omega$ of $\mathcal{P}_n$. For convenience, we denote such stars as $\Omega = \psi \times \pi$. In the plutonium alloy experiment, $PA_2$ satisfies this structure with $\pi = \langle AB, DE, ACD\rangle$ and $\psi=\{\{A\}, \{C\}, \{AC\}\}$. Here, $\psi$ corresponds to a $0$-spread of $\mathcal{P}_2 = \langle A, C \rangle$.

One of the most popular methods of constructing a balanced spread of a finite projective space in $GF(2)$ is the cyclic approach of \cite{hirschfeld1998projective}. 
Suppose $h$ and $u$ are positive integers such that $h$ divides $u$, and we wish to construct an $(h-1)$-spread $\psi=\{f_1,...,f_{\mu}\}$ of $\mathcal{P}_u$, where $\mu=(2^u-1)/(2^h-1)$. 
The cyclic method for constructing $\psi$ starts by writing the $2^u-1$ nonzero elements of $GF(2^u)$ in cycles of length $\mu$ (Table~\ref{tab:cycle}). 
The nonzero elements of $GF(2^u)$ are written as $\left\{\omega^0,\omega^1,\ldots,\omega^{2^u-2}\right\}$, where $\omega$ is a primitive element, and $\omega^i = \alpha_{0}\omega^0+\alpha_{1}\omega^1+\cdots+\alpha_{u-1}\omega^{u-1}$, for $0\le i\le 2^u-2$, correspond to the vector representation $(\alpha_{0},\ldots, \alpha_{u-1})$ of elements in $\mathcal{P}_u$. 
\begin{table}[h!]\centering
\caption{The elements of $GF(2^u)$ in cycles of length $\mu=(2^u-1)/(2^h-1)$.}
\footnotesize{
\begin{tabular}{c c c c}
\hline
$f_{1}$ & $f_{2}$ & $\cdots$ & $f_{\mu}$\\
\hline
$\omega^0$ & $\omega^1$ & $\cdots$ & $\omega^{\mu-1}$\\
$\omega^\mu$ & $\omega^{\mu+1}$ & $\cdots$ & $\omega^{2\mu-1}$\\
$\vdots$ & $\vdots$ & $\ddots$ & $\vdots$ \\
$\omega^{2^u-\mu-1}$ & $\omega^{2^u-\mu}$ & $\cdots$ & $\omega^{2^u-2}$ \\
\hline
\end{tabular}\label{tab:cycle}
}
\end{table}

\cite{hirschfeld1998projective} showed that the $f_i$'s are $(h-1)$-flats and $\psi=\{f_1,...,f_{\mu}\}$ partitions the set of all nonzero elements of $GF(2^u)$, i.e., $\psi$ is an $(h-1)$-spread of $\mathcal{P}_u$. For a quick reference, Table~\ref{tab:spread-example} shows the $2$-spread $\psi$ of $\mathcal{P}_6$ generated using the primitive polynomial $\omega^6 + \omega + 1$ and primitive element $\omega$. Here, $\mu = (2^6-1)/(2^3-1) = 9$ disjoint $2$-flats $\{f_1,..., f_9\}$ can be used to construct different RDCSSs as required. 

\begin{table}[h!]\centering
{
\caption{The $2$-spread obtained using the cyclic construction.}
\footnotesize{
\begin{tabular}{c c c c c c c c c}
\hline
$f_{1}$ & $f_{2}$ &  $f_3$ & $f_4$ & $f_5$ & $f_6$ & $f_7$ & $f_8$ & $f_{9}$\\
\hline
F & E & D & C & B & A & EF & DE & CD \\
BC & AB & AEF & DF & CE & BD & AC & BEF & ADE \\
CDEF & BCDE & ABCD & ABCEF & ABDF & ACF & BF & AE & DEF \\
CDE & BCD & ABC & ABEF & ADF & CF & BE & AD & CEF \\
BDE & ACD & BCEF & ABDE & ACDEF & BCDF & ABCE & ABDEF & ACDF \\
BCF & ABE & ADEF & CDF & BCE & ABD & ACEF & BDF & ACE \\
BDEF & ACDE & BCDEF & ABCDE & ABCDEF & ABCDF & ABCF & ABF & AF \\
\hline
\end{tabular}\label{tab:spread-example}
}}
\end{table}


\section{Isomorphism of RDCSS-based Designs}
In this paper, we only consider the spread- and star-based designs, and for simplicity, we also assume that all RDCSSs of a design are of the same size (so the resulting structure is balanced). 
Let $d_1$ and $d_2$ be two $2^n$ multi-stage factorial designs with $\mu$ stages of randomization, and let the respective RDCSSs be represented by $d_1 = \{f_1,...,f_{\mu}\}$ and $d_2 = \{g_1,...,g_{\mu}\}$.
In spirit of \cite{bingham2008factorial}, the two designs are said to be isomorphic (denoted by $d_1 \cong d_2$) if one can be obtained from the other by applying some sort of relabeling of factors and factor levels and/or reordering of effects within the RDCSSs. 
We formalize this definition by first bundling up the reordering and rearrangement-type operations together in one concept called ``equivalence" and  then addressing the relabeling step.

\begin{definition}\label{def:eqv}
Two $2^n$ RDCSS-based factorial designs $d_1= \{f_1,...,f_{\mu}\}$ and $d_2 = \{g_1,...,g_{\mu}\}$ are said to be \emph{equivalent} (denoted by, $d_1 \equiv d_2$) if and only if, for every $f_i \in d_1$, there is a unique $g_j \in d_2$ such that $\{f_i\} = \{g_j\}$ (set equality), for $1 \le i,j\le \mu$.
\end{definition}

This notion of equivalence will not only take care of rearrangement of factor combinations within a given RDCSS, but also account for reordering of the RDCSSs themselves. Let $\mathcal{E}(d_1)$ be the set of all such $2^n$ designs in $\mu$-stages that are equivalent to $d_1$ (i.e., $\mathcal{E}(d_1)$ denotes the equivalence class of $d_1$). If $|f_i|=2^t-1$ for each $f_i\in d_1$, then,
\begin{equation}\label{eq:equiv_class_size}
|\mathcal{E}(d_1)|= \mu!\cdot [(2^t-1)!]^{\mu}.
\end{equation}

Assuming the existence of a $(t-1)$-spread or a covering star $St(n, \mu, t, t_0)$ of $\mathcal{P}_n$ involved in the construction of the $2^n$ design, the maximum value of $\mu$ is $(2^n-1)/(2^t-1)$ or $(2^{n-t_0}-1)/(2^{t-t_0}-1)$, respectively. 
When $n$ is large, checking the equivalence of two designs by naively iterating through the entire equivalence class of one of them is too computationally intensive, e.g., in the plutonium alloy experiment, $|\mathcal{E}(PA_2)| = 6(15!)^3 \approx 1.37\times 10^{37}$. 
However, the computational burden can be reduced through a combination of sorting and the following \emph{bitstring representation} scheme. 

Each element of $\mathcal{P}_n$ can be represented as a unique binary string of $2^n-1$ bits with exactly one nonzero entry. For instance, following the Yates Order \citep{box1978statistics}  of $\mathcal{P}_3$, the bitstring representations of the elements of $\mathcal{P}_3 = \{A, B, AB, C, AC, BC, ABC\}$ are $A \rightarrow \texttt{1000000}$, $B \rightarrow \texttt{0100000}$, $\ldots$, $ABC \rightarrow \texttt{0000001}$. 
In this representation, the contents of any RDCSS $f$ can now be uniquely identified by the sum of the bitstring representations of its elements. For instance, $f=\{AB, AC, BC\}$ is now uniquely identified by the representation $\texttt{0010110}$. After converting to this representation, checking the equivalence of two RDCSSs amounts to checking the equality of two bitstrings. Furthermore, checking the equivalence of two RDCSS-based designs becomes equivalent to checking the equality of two sets of bitstrings, which is straightforward if one sorts them first. Note that this new bitstring representation is typically more advantageous for spreads with larger values of $t$. For smaller $t$ (say, $2$ or $3$), simply sorting the elements in each RDCSS is sufficient.

We use a collineation of $\mathcal{P}_n$ to express the relabelings of factors and factor combinations \citep{coxeter2003projective}. A \emph{collineation} of $\mathcal{P}_n$ is a mapping of the points from $\mathcal{P}_n$ to $\mathcal{P}_n$ such that $(t-1)$-flats gets mapped to $(t-1)$-flats for all $1\leq t \leq n$. A collineation of $\mathcal{P}_n$ can be characterized by a full rank $n \times n$ matrix $\mathcal{C}$ over $GF(2)$, referred to as the \emph{collineation matrix} \citep{batten1997combinatorics}, where the $j$-th column of $\mathcal{C}$ is the image of the $j$-th basic factor (say) $F_{j}$ (i.e., the factorial effect in $\mathcal{P}_n$ that $F_j$ gets mapped to). {For instance, the $3 \times 3$ collineation matrix 
\[
\mathcal{C} =   \begin{bmatrix} 
      1& 0 & 0\\
      0& 1 & 1\\
      0& 0 & 1\\
   \end{bmatrix}
   \]
relabels the basic factors as $A \rightarrow A$, $B \rightarrow B$ and $C \rightarrow BC$. See Appendix~C for easy implementation in} \verb"R". We interchangeably use the terms ``collineation" and ``collineation matrix" to refer to the same linear mapping. Let $\mathcal{C}_{n}$ be the set of all collineations of $\mathcal{P}_n=PG(n-1,2)$, then the size of $\mathcal{C}_n$ is given by
\begin{equation}\label{eq:coll_size}
|\mathcal{C}_{n}| = \prod_{j=1}^{n} \left(2^n-2^{j-1}\right).
\end{equation} 
The proof of (\ref{eq:coll_size}) follows by simply counting the total number of linearly independent images of $F_j$. That is, given that the images of $F_1,...,F_j$ have already been selected, the total number of possible images for $F_{j+1}$ is $2^{n}-2^{j}$.


Suppose we wish to construct a collineation $\mathcal{C}$ which defines the mapping between $\{x_1, \ldots, x_n\}$ and $\{ y_1, \ldots, y_n \}$, i.e., $\mathcal{C}(x_i) = y_i$, for $1=1,\ldots,n$. One intuitive method of constructing such a collineation matrix is to solve a system of $n$ equations with $n$ unknown over $GF(2)$. As discussed in Algorithm~1 of Section~4, this equation solving approach could be computationally expensive for checking isomorphism. Thus, we propose a two-step alternative approach. First construct a collineation matrix $\mathcal{C}_{x,B}$ that canonicalize $\{x_1, \ldots, x_n\}$, i.e., $x_1 \rightarrow A$, $x_2 \rightarrow B$, and so on, and then construct $\mathcal{C}_{B,y}$ that maps the canonical basis elements to $y_j$'s, i.e., $A \rightarrow y_1$, $B \rightarrow y_2$, etc. As a result the desired collineation matrix is $\mathcal{C} = \mathcal{C}_{B,y}\cdot \mathcal{C}_{x,B}$.

For instance, for the silicon wafers example, let $x_1 = A$, $x_2 = EF$, $x_3 = BCE$, $x_4 = B$, $x_5 = AF$, $x_6 = CDF$ be the effects from RDCSSs in $IC_1$, and $y_1 = A$, $y_2 = BD$, $y_3 = CF$, $y_4 = B$, $y_5 = AF$, $y_6 = CE$ be chosen from RDCSSs in $IC_2$. Then, $\mathcal{C}_{x,B}$ can be constructed by writing $x_i$'s as column vectors and then inverting the matrix, i.e., 
\[
\mathcal{C}_{x,B} =   \begin{bmatrix} 
      1& 0 & 0& 0 & 1 & 0\\
      0& 0 & 1& 1 & 0 & 0\\
      0& 0 & 1& 0 & 0 & 1\\
      0& 0 & 0& 0 & 0 & 1\\
      0& 1 & 1& 0 & 0 & 0\\
      0& 1 & 0& 0 & 1 & 1\\
   \end{bmatrix}^{-1}
    =  \begin{bmatrix} 
      1& 0 & 1& 0 & 1 & 1\\
      0& 0 & 1& 1 & 1 & 0\\
      0& 0 & 1& 1 & 0 & 0\\
      0& 1 & 1& 1 & 0 & 0\\
      0& 0 & 1& 0 & 1 & 1\\
      0& 0 & 0& 1 & 0 & 0\\
   \end{bmatrix}.
   \]

From an implementation standpoint the inversion of a matrix $M$ over $GF(2)$ can easily be done in R \citep{R-Core-Team:2014aa} using the code: \verb"solve(M)%%2".
As highlighted in Algorithm~1, this inversion is required much less often as compared to the naive method (via solving a system of $n$ equations in $n$ unknowns). Next, $\mathcal{C}_{B,y}$ is constructed by simply writing $y_j$'s as column vectors. As a result, the desired collineation matrix that defines the mapping from $x_i$'s to $y_j$'s is given by
\[
\mathcal{C} = \mathcal{C}_{B,y}\cdot\mathcal{C}_{x,B} =   \begin{bmatrix} %
      1& 0 & 0 & 0 & 1 & 0\\
      0& 1 & 0 & 1 & 0 & 0\\
      0& 0 & 1 & 0 & 0 & 1\\
      0& 1 & 0 & 0 & 0 & 0\\
      0& 0 & 0 & 0 & 0 & 1\\
      0& 0 & 1 & 0 & 1 & 0\\
   \end{bmatrix} 
    \cdot 
    \begin{bmatrix} 
      1& 0 & 1& 0 & 1 & 1\\
      0& 0 & 1& 1 & 1 & 0\\
      0& 0 & 1& 1 & 0 & 0\\
      0& 1 & 1& 1 & 0 & 0\\
      0& 0 & 1& 0 & 1 & 1\\
      0& 0 & 0& 1 & 0 & 0\\
   \end{bmatrix}
   = 
    \begin{bmatrix} %
      1& 0 & 0& 0 & 0 & 0\\
      0& 1 & 0& 0 & 1 & 0\\
      0& 0 & 1& 0 & 0 & 0\\
      0& 0 & 1& 1 & 1 & 0\\
      0& 0 & 0& 1 & 0 & 0\\
      0& 0 & 0& 1 & 1 & 1\\
   \end{bmatrix}.
   \]

Applying this collineation on RDCSSs of $IC_1$ gives $\mathcal{C}(IC_1) = \{ \langle A, BD, EF\rangle$, $\langle B, AF, CE\rangle$, $\langle CD, AB, ABE\rangle$, $\langle DEF, BCD, D\rangle$, $\langle BDF, CEF, ACDF\rangle$, $\langle F, BE, ABDEF\rangle$, $\langle BDEF, BF,$ $ABCF\rangle$, $\langle ACD, BCF, BDE\rangle$, $\langle ADEF, DF, CDF\rangle \}$.

We now combine the notion of equivalence and collineation to formally define the isomorphism of multi-stage factorial designs characterized by the RDCSSs.

\begin{definition}\label{def:iso}
Two $2^n$ RDCSS-based factorial designs $d_1= \{f_1,...,f_{\mu}\}$ and $d_2 = \{g_1,...,g_{\mu}\}$ are said to be \emph{isomorphic} (denoted by $d_1 \cong d_2$) if and only if there exists a collineation $\mathcal{C}$ over $\mathcal{P}_n$ such that $\mathcal{C}(d_1) \equiv d_2$. In this case, we say that $\mathcal{C}$ is an isomorphism establishing collineation (IEC) from $d_1$ to $d_2$.
\end{definition}

For any pair of RDCSS-based designs in $\mathcal{P}_n$, if there exists one IEC, then there are typically many more - obtained by multiplying with the collineation matrices in $\mathcal{C}_n$. However, determining if one IEC exists is the most difficult part. For instance, in the silicon wafers example, $\mathcal{C}(IC_1)$ (discussed above) is not equivalent to $IC_2$. This is apparent when comparing their bitstring representations. Let $a_{(1)}, \ldots, a_{(9)} \in \left\{0,1\right\}^{63}$ denote the bitstring representations of the RDCSSs of $\mathcal{C}(IC_1)$ (sorted according to their first nonzero entries) and let $b_{(1)}, \ldots, b_{(9)} \in \left\{0,1\right\}^{63}$ denote the analogous sorted representation for $IC_2$. Note that each bitstring, here, would be of length 63 with only seven 1's and the rest zeros. Determining these bitstring representations reveals that
\begin{align*}
a_{(1)} &= \texttt{100000000110000000000000000000000000000000000001100000000110000},\\
b_{(1)} &= \texttt{100000000110000000000000000000000001100000000001100000000000000}.
\end{align*}
Clearly, $a_{(1)} \neq b_{(1)}$, which is sufficient to conclude $\mathcal{C}(IC_1) \not\equiv IC_2$. Therefore, $\mathcal{C}$ is not an IEC from $IC_1$ to $IC_2$. However, all viable collineations must be checked before we could conclude that $IC_1$ and $IC_2$ are non-isomorphic. If we assume that $IC_1$ and $IC_2$ are isomorphic, how do we find an IEC? 

Note that the results discussed thus far in this section do not assume any specific  overlapping pattern of the RDCSSs. Next, we discuss the IEC search algorithms separately for spread-based designs (with disjoint RDCSSs), and the star-based designs, where all RDCSSs have a common overlap.


\section{Search Algorithm for Spread-based Designs}

When undertaken naively, determining if two RDCSS-based designs on $\mathcal{P}_n$ are isomorphic involves exhaustively searching over the entire space $\mathcal{C}_n$--- as enumerated by (\ref{eq:coll_size})--- to check if any are IECs. 
Depending on the value of $n$, this full search space can be prohibitively large, meaning an exhaustive search would be computationally intractable. 
We propose a strategy that exploits the structure of IECs to reduce the size of this search space. 
Here, we motivate and describe our strategy as it applies to RDCSS-based designs with no overlap. 
In Section~5, we extend this approach to RDCSSs that share a common overlap (i.e., star-based designs).

From the definition of isomorphism (Definition~\ref{def:iso}), a collineation $\mathcal{C}$ can be considered as an IEC if and only if, for all $i \in \{1, 2, ..., \mu\}$, the RDCSS $f_i$ in $d_1$ is relabeled by $\mathcal{C}$ to $g_j$ in $d_2$ for some $j \in \{1, 2, ..., \mu\}$. 
We refer to this as the \emph{full RDCSS mapping property} (FRMP) of an IEC. 
The FRMP characterizes which collineations to consider when searching for an IEC. 
For example, if $\mathcal{C}$ is an IEC from $IC_1$ to $IC_2$ for the silicon wafers example in Section~2.1, then we know that $\mathcal{C}(A)$, $\mathcal{C}(EF)$, and $\mathcal{C}(BCE)$ must all fall within the same RDCSS of $IC_2$; 
we need not consider any $\mathcal{C}$ for which $\mathcal{C}(A)$ and $\mathcal{C}(EF)$ belong to different RDCSSs, such as  $\mathcal{C}(A)= B $ and $\mathcal{C}(EF)= C$. 
Furthermore, because $A$ and $B$ are in different RDCSSs in $IC_1$, we know that $A$ and $B$ must be mapped to different RDCSSs in $IC_2$; we can rule out any other collineation, such as those for which $A \rightarrow BD$ and $B \rightarrow A$, when searching for an IEC.  

Unfortunately, we are not aware of any constructive approaches for isolating collineations that completely satisfy the FRMP. For this reason, we introduce a relaxed version of the FRMP --- called the \emph{partial RDCSS mapping property} (PRMP) --- that does admit a constructive approach. A collineation $\mathcal{C}$ satisfies the PRMP for points $x_1, \ldots, x_n \in \mathcal{P}_n$ if it meets the following requirement: for all $i,j: 1\leq i,j \leq n$, there exist unique $k, l\ (1\le k \ne l\le \mu)$ such that, $x_i,x_j \in f_k$ if and only if $\mathcal{C}(x_i), \mathcal{C}(x_j) \in g_{\ell}$.  That is, a collineation $\mathcal{C}$ satisfies the PRMP for $x_1,\ldots, x_n$ if $x_i,x_j$ being co-members of an RDCSS in $d_1$ occurs if and only if $\mathcal{C}(x_i)$ and $\mathcal{C}(x_j)$ are co-members of an RDCSS in $d_2$.

Our reasons for defining the PRMP in this manner are two-fold. Firstly, the constraints imposed by the PRMP are a strict subset of the constraints imposed by the FRMP. As a result, any collineation satisfying the FRMP (i.e. IECs) will also satisfy the PRMP for any given $x_1, \ldots, x_n$; our reduction of the search space does not ignore any possible IECs. Secondly, because the constraints in the PRMP involve just $n$ points, the collineations satisfying the PRMP are now straightforward to construct. 

Recall that for linearly independent $x_1,\ldots, x_n$, any collineation $\mathcal{C}$ is characterized by the images $y_1 =\mathcal{C}(x_1),\ldots, y_n = \mathcal{C}(x_n)$ with the corresponding collineation matrix being easily determined using the algorithm provided in Section~3. 
Thus, the entire search space for IECs given by a PRMP can be constructed by iterating through the possible options for $y_1, \ldots, y_n$ in $d_2$ that match the RDCSS co-membership structure of $x_1,\ldots, x_n$ in $d_1$. 
The remainder of this section describes an approach that efficiently iterates through the collineation matrices in this class to look for IECs. 
We can then check the equivalence of $\mathcal{C}(d_1)$ and $d_2$ --- via the bitstring representation --- for every $\mathcal{C}$ in this class. 

We now develop some theory to provide guidance on how to choose the basis set $x_1,\ldots, x_n \in \mathcal{P}_n$. Proposition~\ref{propsize} presents an upper bound on the number of collineations satisfying the PRMP for a basis $\{x_1,\ldots, x_n\}$. Appendix~A provides a proof of this result, as well as a comment explaining why it is only an upper bound.

\begin{prop}\label{propsize}
Let $d_1 = \{f_1,\ldots, f_{\mu}\}$ and $d_2 = \{g_1, \ldots, g_{\mu}\}$ be two $2^n$ RDCSS-based designs obtained from balanced $(t-1)$-spreads of $\mathcal{P}_n$. Let $\{x_1,\ldots, x_n\}$ be a basis of $\mathcal{P}_n$, $m_i = |f_i \cap \left\{x_1, \ldots, x_n \right\}|$ for $i=1,\ldots, \mu$, and $\ell$ be the number of nonzero $m_i$'s. 
Then, the number of collineations $\mathcal{C}$ from $d_1$ to $d_2$ which satisfy the PRMP for $x_1, \ldots, x_n$ is bounded above by
\begin{equation}\label{count}
\frac{\mu!}{(\mu-\ell)!}  \prod_{i=1}^{\ell}\left(\prod_{j=1}^{m_i} \left( 2^t-2^{j-1} \right)\right).
\end{equation}
\end{prop}

Recall that a balanced $(t-1)$-spread of $\mathcal{P}_n$ exists if and only if $t$ divides $n$. Therefore, the upper bound (\ref{count}) from Proposition~\ref{propsize} is minimized for $\ell = n/t$, which further implies either $m_i= t$ or zero. Proposition~\ref{itspossible} guarantees the existence of such a basis for any set of RDCSSs obtained from a balanced $(t-1)$-spread of $\mathcal{P}_n$ (see Appendix~A for the proof).

\begin{prop}\label{itspossible}
For any balanced $(t-1)$-spread based multi-stage design $d_1=\{f_1, \ldots, f_{\mu}\}$ in $\mathcal{P}_n$, there exists $\ell_0=n/t$ distinct RDCSSs $f_{u_1}, \ldots, f_{u_{\ell_0}}$ from $d_1$, with $1\le u_1, \ldots, u_{\ell_0}\le \mu$, such that $\langle \cup_{i=1}^{\ell_0} f_{u_i} \rangle = \mathcal{P}_n$. 
\end{prop}

Subsequently, there exist a set $x_1,\ldots, x_n$ such that the maximum number of collineations that satisfy PRMP can be reduced to
\begin{align}\label{finalcount}
\frac{\mu!}{(\mu - n/t)!} \left(\prod_{j=1}^{t} \left( 2^t-2^{j-1} \right)\right)^{n/t}.
\end{align}
As compared to the naive approach (enumerated in (\ref{eq:coll_size})), the proposed approach corresponds to a reduction of the search space by 7 orders of magnitude for checking the isomorphism of 2-spreads of $\mathcal{P}_6$. 
For 1-spreads and 4-spreads of $\mathcal{P}_{10}$, the search space is reduced by 13 orders and 12 orders of magnitude, respectively. For larger $n$, the improvements are even greater.


%
For checking the isomorphism of $d_1=\{f_1,\ldots, f_{\mu}\}$ and $d_2=\{g_1,\ldots, g_{\mu}\}$ in $\mathcal{P}_n$, we follow a systematic approach to search for the IEC by iterating through the candidate collineations. First, choose $\ell_0=n/t$ out of $\mu$ RDCSSs in $d_1$ that can generate a basis $\{x_1,\ldots,x_n\}$ for $\mathcal{P}_n$ such that each of the $\ell_0$ RDCSSs contributed $t$ points to the basis (as in Proposition~\ref{itspossible}). Then, construct a collineation matrix $\mathcal{C}_{x,B}$ to transform the $x$'s to a canonical basis with basic factors, $\{A, B, ...\}$ (as demonstrated in Section~3). Note that the isomorphism of $\mathcal{C}_{x,B}(d_1)$ and $d_2$ implies the isomorphism of $d_1$ and $d_2$. Now, iterate through all possible sets of $\ell_0$  RDCSSs from $d_2$ to construct the basis set $\{y_1, \ldots, y_n\}$ in $d_2$ and the corresponding collineation matrix $\mathcal{C}_{B,y}$ that maps the canonical basis to $y$'s. If $\mathcal{C}_{B,y}(\mathcal{C}_{x,B}(d_1))$ is equivalent to $d_2$, then we have found the IEC. The step-by-step algorithm is summarized in Algorithm~1. Note that two subscripts for the $x$'s and $y$'s are introduced to keep track of their RDCSS membership in $d_1$ and $d_2$, respectively. Let $F_1, \ldots F_n$ be an alternative notation for the basic factors, with the assumption that $F_1 := A$, $F_2 := B$, and so on. 

\begin{algorithm}[h!]%
\caption{Isomorphism check between two $(t-1)$-spread based designs in $\mathcal{P}_n$.}
\begin{enumerate}
\item Choose $\ell_0=n/t$ out of $\mu$ RDCSSs from $d_1=\{f_1,\ldots, f_{\mu}\}$, that satisfy Proposition~2. Let these be $\{f_{u_1}, \ldots, f_{u_{\ell_0}}\}$ for $1\le u_1, \ldots, u_{\ell_0}\le \mu$.

\item  For $i=1,\ldots, \ell_0$, specify $\{x_{i,1},\ldots,x_{i,t}\} \in f_{u_i}$ such that $f_{u_i} = \langle x_{i,1},\ldots,x_{i,t} \rangle$. 

\item Construct the collineation matrix $\mathcal{C}_{x,B}$ which maps the $x$'s in Step 2 to the canonical basis $\{A, B, \ldots\}$ such that each $x_{i,j}$ is mapped to $F_{t(i-1) + j}$. 

\item
	
	\begin{enumerate}
		\item Choose $\ell_0$ out of $\mu$ RDCSSs from $d_2=\{g_1,\ldots, g_{\mu}\}$ for mapping $f_{u_i}$'s (note that ordering is important). Let that be $\{g_{v_1}, \ldots, g_{v_{\ell_0}}\}$ for $1\le v_1, \ldots, v_{\ell_0}\le \mu$. If $|\langle g_{v_1} \cup \ldots \cup g_{v_{\ell_0}} \rangle| <  2^n-1$, proceed to the next choice for $\{g_{v_1}, \ldots, g_{v_{\ell_0}}\}$.
		
		\item  For $i=1,\ldots, \ell_0$, choose a $\{y_{i,1},\ldots y_{i,t}\} \in g_{u_i}$ such that $g_{v_i} = \langle y_{i,1},\ldots,y_{i,t} \rangle$.
		\item Choose one of  the $(\ell_0)!$ permutations of the elements $1,\ldots, \ell_0$, say $\sigma_k$, for $k = 1, \ldots, (\ell_0)!$.		
		\item Construct $\mathcal{C}_{B,y}$ which maps the canonical basis elements to $y$'s such that $F_{t(i-1) + j}$ is mapped to $y_{\sigma_k(i),j}$.	
		\item If $\mathcal{C}_{B,y}(\mathcal{C}_{x,B}(d_1))$ is equivalent to $d_2$ (as per the bitstring method in Section~3), then, $d_1\cong d_2$, and report $\mathcal{C} = \mathcal{C}_{B,y}\cdot\mathcal{C}_{x,B}$ as an IEC and exit; otherwise, continue.
		\item Go to Step~4(c) and choose another ordering $\sigma_k$ if possible, otherwise, continue.
		\item Go to Step~4(b) and choose another basis if possible, otherwise, continue.
		\item Go to Step~4(a) and choose another set of RDCSSs if possible, otherwise report that $d_1$ and $d_2$ are non-isomorphic.				
	\end{enumerate}	
	
\end{enumerate}	 
\end{algorithm}

%
A few quick remarks worth noting. From an implementation standpoint, the stages in Step~4 are nested, making it straightforward to parallelize at any stage of the hierarchy. The matrix inversion is required only for Step~3, and the factorization $\mathcal{C} = \mathcal{C}_{B,y}\cdot\mathcal{C}_{x,B}$ avoids the need to repeatedly solve that systems of linear equations associated with $\mathcal{C}(x_{i,j}) = y_{\sigma_k(i),j}$. Moreover, Algorithm 1 iterates through all possible collineations satisfying the PRMP for a particular chosen set $\{x_1,\ldots, x_n\}$. As discussed above, the set of collineations satisfying a PRMP is a superset of the collineations satisfying the FRMP (which contains all possible IECs). Therefore, by exhausting the set of collineations for one chosen set $\{x_1,\ldots, x_n\}$, we are guaranteed to have already visited all possible IECs. 

{In terms of algorithmic complexity, the speed-up of moving from the naive search method to Algorithm~1 is proportional to the reduction in the search space size from (\ref{eq:coll_size}) to (\ref{finalcount}). The complexity of the equivalence check algorithm (as described in Section~3) is $O(n^2 2^n)$.} {We have also implemented this algorithm in} \verb"R" (see Appendix~C for illustration).

\textbf{Example:} To demonstrate Algorithm~1, we will illustrate a run to check the isomorphism of $IC_1$ and $IC_2$ (in the silicon wafers experiment) below. Recall that $IC_1 = \{f_1 = \langle A, EF, BCE\rangle$, $f_2 = \langle B, AF, CDF \rangle$, $f_3 =\langle C, AB, ADE\rangle$, $f_4 =\langle D, BC, BEF\rangle$, $f_5 =\langle E, CD, ACF\rangle$, $f_6 =\langle F, DE, ABD\rangle$, $f_7 =\langle BD, BF, ACE\rangle$, $f_8 =\langle AC, CE, BDF\rangle$, $f_9 =\langle AD, BE, CF\rangle \}$, and $IC_2 = \{g_1 =\langle A,BD,CF\rangle$, $g_2 = \langle B,AF,CE\rangle$, $g_3 = \langle C,BF,DE\rangle$, $g_3 = \langle D,AC,BE \rangle$, $g_5 = \langle E,AB,DF\rangle$, $g_6 = \langle F, AE, CD\rangle$, $ g_7 = \langle AD,BC,EF\rangle$, $g_8 = \langle ACE,ADF,BEF\rangle$, $g_9 = \langle ABC,ADE,CEF\rangle\}$. The following represents an iteration of Step 4 for which an ICE from $IC_1$ to $IC_2$ is found.

\begin{enumerate}
\item Here, $\ell_0=2$ with $f_{u_1} = \langle A, EF, BCE\rangle$ and $f_{u_2} =\langle B, AF, CDF \rangle$.

\item  Let $x_{1,1} = A, x_{1,2} = EF, x_{1,3} = BCE$, and  $x_{2,1} = B, x_{2,2} = AF, x_{2,3} = CDF$.

\item The collineation matrix $\mathcal{C}_{x,B}$ that canonicalize these $x$'s has been presented in Section~3.

\item Here, we calculate the total number of options at each stage, and then demonstrate their values when the first IEC found.
	\begin{enumerate}
		\item There are $9!/(9-2)! = 72$ choices for $\{v_1,v_2\} \subset \left\{1,\ldots, 9\right\}$. Our first choice of $\{v_1,v_2\} = \{1,2\}$ leads to an IEC.
		
		\item There are $ (2^3-1)\cdot (2^3-2)\cdot(2^3-2^2) = 7 \cdot 6 \cdot 4 = 168$ choices for linearly independent $y_{1,1}, y_{1,2},y_{1,3} \in g_{1}$, and 168 choices for linearly independent $y_{2,1}, y_{2,2},y_{2,3} \in g_{2}$. Out of $(168)^2$ iterations, we found an IEC in the 359th step of our search. The $y$'s that led to the first IEC are $y_{1,1} = A$, $y_{1,2} = BD$, $y_{1,3} = BCDF$, $y_{2,1} = B$, $y_{2,2} = ABCEF$, $y_{2,3} = ABF$.
		\item There are two options for the permutation: $\sigma_1 = (1,2)$ or $\sigma_2 = (2,1)$. Our first IEC was found using $\sigma_1$.
		\item Here, 
		\[
\mathcal{C}_{B,y} =   \begin{bmatrix} 
      1& 0 & 0& 0 & 1 & 1\\
      0& 1 & 1& 1 & 1 & 1\\
      0& 0 & 1& 0 & 1 & 0\\
      0& 1 & 1& 0 & 0 & 0\\
      0& 0 & 0& 0 & 1 & 0\\
      0& 0 & 1& 0 & 1 & 1\\
   \end{bmatrix}.	
   \]
		\item A run of the equivalence check algorithm verifies that $\mathcal{C}_{B,y}(\mathcal{C}_{x,B}(d_1))$ is equivalent to $d_2$ (details omitted for space). Thus, $d_1\cong d_2$, and 
		\[
		\mathcal{C} = \mathcal{C}_{B,y}\cdot\mathcal{C}_{x,B}=
		\begin{bmatrix} 
      1& 0 & 0& 0 & 1 & 1\\
      0& 1 & 1& 1 & 1 & 1\\
      0& 0 & 1& 0 & 1 & 0\\
      0& 1 & 1& 0 & 0 & 0\\
      0& 0 & 0& 0 & 1 & 0\\
      0& 0 & 1& 0 & 1 & 1\\
   \end{bmatrix}			
    \begin{bmatrix} 
      1& 0 & 1& 0 & 1 & 1\\
      0& 0 & 1& 1 & 1 & 0\\
      0& 0 & 1& 1 & 0 & 0\\
      0& 1 & 1& 1 & 0 & 0\\
      0& 0 & 1& 0 & 1 & 1\\
      0& 0 & 0& 1 & 0 & 0\\
   \end{bmatrix}
   =
       \begin{bmatrix} 
      1& 0 & 0& 1 & 0 & 0\\
      0& 1 & 0& 0 & 0 & 1\\
      0& 0 & 0& 1 & 1 & 1\\
      0& 0 & 0& 0 & 1 & 0\\
      0& 0 & 1& 0 & 1 & 1\\
      0& 0 & 0& 0 & 1 & 1\\
   \end{bmatrix}
   \]
	is an IEC. We can exit the algorithm.		
	\end{enumerate}	
	
\end{enumerate}		

\section{Search Algorithm for Star-based Designs}

Recall that a balanced star-based design refers to a multi-stage factorial design with equal sized RDCSSs that share a common overlap. As per Lemma~1, a star $\Omega =St(n, \mu, t, t_0)$ can be expressed as $\Omega = \psi \times \pi$, where $\psi$ is a $((t-t_0) -1)$-spread of $\mathcal{P}_{n-t_0}$, and $\pi$ is $(t_0-1)$-dimensional subspace in $\mathcal{P}_n \backslash \mathcal{P}_{n-t_0}$. As earlier, let $u = n-t_0$ and $h=t-t_0$. Thus, the isomorphism check between two star-based designs $d_1$ and $d_2$ (with $(t_0-1)$-dimensional nuclei) can be reduced to checking isomorphism between two $(h-1)$-spreads of $\mathcal{P}_{n-t_0}$ by iterating through the elements of  $\mathcal{C}_{n - t_0}$ instead of  $\mathcal{C}_{n}$. Even for small nuclei (e.g. $t_0 =1$ or $2$), this corresponds to a large reduction in the search space.

Algorithm~2 summarizes the steps of how Algorithm~1 can be used to search for an IEC between $d_1$ and $d_2$ based on $\Omega_1 = \psi_1 \times \pi_1$ and $\Omega_2 = \psi_2 \times \pi_2$, respectively.

\begin{algorithm}[h!]%
\caption{Isomorphism check between two $St(n, \mu, t, t_0)$-based designs $d_1$ and $d_2$, which correspond to stars $\Omega_1 = \psi_1 \times \pi_1$ and $\Omega_2 = \psi_2 \times \pi_2$, respectively.}
\begin{enumerate}
\item Determine two bases $\{p_{1,1},\ldots, p_{1,t_0}\}$ and $\{p_{2,1}, \ldots, p_{2,t_0}\}$ of the nuclei $\pi_1$ and $\pi_2$, respectively. 

\item Construct a collineation matrix $\mathcal{C}_{\pi_1, B}$ mapping $\{p_{1,1},\ldots, p_{1,t_0}\}$ to the $t_0$ trailing basic factors $F_{n - t_0 + 1}, \ldots, F_{n}$. The pre-images of $F_1, \ldots, F_{n-t_0}$ can be chosen as an arbitrary linearly independent set from $\mathcal{P}_n \backslash \pi_1$.

\item Similarly, construct a collineation matrix $\mathcal{C}_{\pi_2, B}$ mapping $\{p_{2,1}, \ldots, p_{2,t_0}\}$ to the $t_0$ trailing basic factors $F_{n - t_0 + 1}, \ldots, F_{n}$.

\item Extract designs $d_1^*$ and $d_2^*$ on $\mathcal{P}_{n-t_0}$ corresponding to the spreads $\mathcal{C}_{\pi_1, B}(\psi_1)$ and $\mathcal{C}_{\pi_2, B}(\psi_2)$.

\item Run Algorithm 1 on $d_1^*$ and $d_2^*$. If we come across a $\mathcal{C}^*$ that is an IEC, then an IEC for $d_1$ and $d_2$ is given by $\mathcal{C} = \mathcal{C}_{\pi_2, B}^{-1} \cdot \mathcal{C}^* \cdot \mathcal{C}_{\pi_1, B}$. Otherwise, $d_1$ and $d_2$ are non-isomorphic.

\end{enumerate}
\end{algorithm}

Note that in Algorithm~2, Steps 1-4 relabel the $(h-1)$-spreads $\psi_1$ and $\psi_2$ of $\mathcal{P}_{n-t_0}$ such that their points are within the span of the first $n-t_0$ canonical factors $F_1, \ldots, F_{n-t_0}$ (or $A, B, ...$). This relabeling allows Algorithm~1 to be run using a search space of collineations over $\mathcal{P}_{n-t_0}$ rather than over $\mathcal{P}_{n}$, as all effects involving the last $t_0$ basic factors are discarded along with the nuclei. This reduction of the search space decreases both the number of collineations that need to be considered as well as the dimension of the structures involved.

{Similar to Algorithm 1, the computational complexity of Algorithm~2 improves over the naive method proportionally to the reduction in the search space from (\ref{eq:coll_size}) evaluated at $n$ to (\ref{finalcount}) evaluated at $n-t_0$. Furthermore, the complexity of each equivalence check decreases from $O(n^2 2^n)$ to $O(n^2 2^{n-t_0})$ because the objects being compared are smaller after reducing to spreads.} {Appendix~C illustrates the usage of our} \verb"R" implementation.

In the plutonium alloy example, $PA_1=\{\langle A, B, CDE\rangle$, $\langle C, AD, BE\rangle$, $\langle D, E, ABC\rangle\}$ is derived from a balanced covering star $St(5, 5, 3, 1)$ of $\mathcal{P}_5$, whereas, $PA_2 = \{\langle A, B, DE, ACD\rangle$, $\langle C, AB, DE, ACD\rangle$, $\langle D, E, AB, ACD\rangle\}$ represents a covering star $St(5, 3, 4, 3)$ of $\mathcal{P}_5$. Since the sizes of the two nuclei are different, the two designs are trivially non-isomorphic.

\section{Results on Special Cases}

Thus far we have developed theoretical results and algorithms for checking whether or not two RDCSS-based designs are isomorphic. However, we often want to compare all possible admissible designs and find the optimal one as per some ranking criterion. For this purpose, we need to construct all possible different (or non-isomorphic) admissible designs. 
This is a much bigger challenge for RDCSS-based designs because formal construction methods for all possible spreads or stars are not known yet. 
In this section, we discuss some results from the Combinatorics literature on the complete classification of spreads that can be used for small RDCSS-based designs in $\mathcal{P}_n$.

\subsection{Complete Classification}\label{classification}

A balanced $(t-1)$-spread of  $\mathcal{P}_n$ is referred to as \emph{trivial} for $t=1$ and $t=n$. Of course, we are more interested in the non-trivial cases. As expected, complete classification of non-isomorphic balanced spreads is known for only small $n$.

\begin{enumerate}[{{\; A}1.}]
\item $[n\le 3]$: All balanced spreads are trivial, meaning for every given $n\le 3$ all $(t-1)$-spreads of $\mathcal{P}_n$ are isomorphic. 

\item $[n=4]$: \cite{andre1954nicht} ensures the existence of non-trivial spreads for $t=2$. \cite{soicher2000computation} show that all $1$-spreads of $\mathcal{P}_4$ are isomorphic.

\item $[n=5]$: All covering spreads are trivial, as $n=5$ is prime. However, a complete classification of partial spreads of $\mathcal{P}_5$ is available in \cite{gordon2004classification}, indicating that there are 4 isomorphism classes of maximal $1$-spreads of $\mathcal{P}_5$ consisting of 9 RDCSSs.

\item $[n=6]$: There exist non-trivial balanced spreads for $t=2$ and $t=3$. \cite{topalova20102} showed that all $2$-spreads ($t=3$) of $\mathcal{P}_6$ are isomorphic, and \cite{mateva2009line} used exhaustive search to show that there exist 131044 mutually non-isomorphic $1$-spreads ($t=2$) of $\mathcal{P}_6$.

\item $[n=7]$: All covering spreads are trivial, as $n=7$ is prime. However, \cite{honold2019classification} have shown that there exist 715 isomorphism classes of maximal partial $2$-spreads of $\mathcal{P}_7$ consisting of 17 RDCSSs.
\end{enumerate}

For $n\ge 8$, we are unaware of any such results. For an example of two non-isomorphic $1$-spreads of $\mathcal{P}_6$, consider $d_1 = \{f_1,...,f_{21}\}$, shown in Table~\ref{tab:nonisospreads}, and $d_2 =\{g_1,...,g_{21}\}$, where $g_i=f_i$ for $i=1,...,18$, and $g_{19} = \{ACE, AF,CEF\}$, $g_{20} = \{BCDF,CF,BD\}$, and $g_{21} = \{ABCD,AEF,BCDEF\}$. Here, the spread for $d_2$ was obtained via partitioning the RDCSSs $f_{19}$, $f_{20}$, and $f_{21}$ in $d_1$ into 3 new RDCSSs. The non-isomorphism of the spreads was verified using Algorithm 1.

\begin{table}[h!]\centering
\caption{A $1$-spread of $\mathcal{P}_6=PG(5,2)$}
\label{tab:nonisospreads}
\scriptsize{
\begin{tabular}{c c c c c c c c c c c}
\hline
$f_{1}$ & $f_{2}$ &	$f_{3}$ & $f_{4}$ & $f_{5}$ & $f_{6}$ & $f_{7}$ &	$f_{8}$\\
\hline
F & E & D & C & B & A & EF & DE\\
ABCEF & ABDF & ACF & BF & AE & DEF & CDE & BCD\\
ABCE & ABDEF & ACDF & BCF & ABE & ADEF & CDF & BCE \\
\hline
$f_{9}$ & $f_{10}$& $f_{11}$ & $f_{12}$ &	$f_{13}$ & $f_{14}$ & $f_{15}$ & $f_{16}$ \\
 \hline
CD & BC & AB  & DF & CE  & AC & BEF & ADE \\
ABC & ABEF & ADF  & BE & AD  & BDE & ACD & BCEF \\
ABD & ACEF & BDF  & BDEF & ACDE & ABCDE & ABCDEF & ABCDF \\
\hline
$f_{17}$ &	$f_{18}$ & $f_{19}$ & $f_{20}$& $f_{21}$\\
\hline
 CDEF & BCDE & ABCD& AEF& BD \\
ABDE & ACDEF & BCDF& CF & CEF \\
 ABCF & ABF & AF& ACE & BCDEF \\
\hline
\end{tabular}\label{tab:nonisospreads}
}
\end{table}

Admittedly, multi-stage factorial experiments with a small number of basic factors have been typically more common in industrial experiments. 
However some of the modern experiments, for instance using computer simulation models, involve a large number of factors. 
We now argue that the classification of balanced spreads presented above generates a rich class of star-based designs. 

One may not find $0$-spreads to be useful for a spread-based design, but such spreads can play a crucial role in constructing useful star-based designs, e.g., via  Lemma~\ref{lemma:starspreadcorr}. For instance, in the plutonium alloy example, a $0$-spread of $\mathcal{P}_2$ is used to construct the balanced covering star $St(5,3,4,3)$ of $\mathcal{P}_5$ which generates $PA_2$ -- a 3-stage $2^5$ split-lot design with four randomization factors at each stage. 
This design is desirable as the significance of all effects can be assessed using four half-normal plots (see \cite{ranjan2010stars} for details).  Similarly, $1$-spreads and $2$-spreads can also be augmented with different sized nuclei to form a variety of stars, providing flexibility for constructing small to large star-based designs. 

We now use the results from A1 -- A5 and the theoretical results presented in Sections~3 and 5 to classify the non-isomorphic balanced covering stars $St(n, \mu, t, t_0)$ of $\mathcal{P}_n$. For convenience we follow the same notation, $u=n-t_0$ and $h=t-t_0$, as in Lemma~\ref{lemma:starspreadcorr}.

\begin{enumerate}[{{\; B}1.}]
\item For any given $0\le u \le 5$, $t_0 \ge 0$ and $h$ that divides $u$, all balanced covering stars $St(u+t_0, \mu, h+t_0, t_0)$ of $\mathcal{P}_{u+t_0}$ are isomorphic to each other. The proof follows from Lemma~1 and A1 -- A3.

\item For $t_0 \ge 0$, all balanced covering stars $St(6+t_0, \mu, 2+t_0, t_0)$ of $\mathcal{P}_{6+t_0}$ are isomorphic to each other. The result follows from A4.

\item For every $t_0 \ge 0$, there exist 131044 mutually non-isomorphic balanced stars $St(6+t_0, \mu, 3+t_0, t_0)$ of $\mathcal{P}_{6+t_0}$. The result follows from A4.
\end{enumerate}

The categories defined by B1 and B2 contain most of the popular stars used to obtain RDCSS-based designs. For cases falling outside of these categories, it may be necessary to search over representatives from all isomorphism classes when searching for designs. 
This is a difficult problem because other than the exhaustive search methods used by \cite{topalova20102} and \cite{mateva2009line}, there is no known strategy for obtaining representatives of all isomorphism classes. 


\subsection{Cyclic-spread based Designs}\label{cyclic}

In this section we provide results demonstrating that the most popular approach of constructing a balanced spread -- the cyclic approach of \cite{hirschfeld1998projective}  -- only accesses a single isomorphism class, even if many exist. 
The algebraic results used in the proofs (presented in Appendix~B) are mostly based on \cite{lidl1994introduction}.

Although the spreads obtained via the cyclic construction method (outlined in Section~2.2) may vary with the choice of the primitive element and primitive polynomial of $GF(2^u)$, the next two results establish that such spreads are equivalent or isomorphic.

\begin{theorem}\label{thm:cyclicequiv}
Let $\psi_{1}=\{f_1,...,f_{\mu}\}$ and $\psi_{2}=\{g_1,...,g_{\mu}\}$ be two $(h-1)$-spreads of $\mathcal{P}_u$ constructed using the cyclic method with two different roots $\alpha$ and $\beta$ of the same primitive polynomial $P(\omega)$. Then $\psi_{1}$ is equivalent to $\psi_{2}$.
\end{theorem}

\begin{theorem} \label{thm:cyclic_iso}
Let $\psi_{1}=\{f_1,...,f_{\mu}\}$ and $\psi_{2}=\{g_1,...,g_{\mu}\}$ be two $(h-1)$-spreads of $\mathcal{P}_u$ constructed using the cyclic method with two different primitive polynomials $P_1(\omega)$ and $P_2(\omega)$ respectively. Then $\psi_{1}$ is isomorphic to $\psi_{2}$.
\end{theorem}

See Appendix~B for the proofs of Theorems~1 and 2. Although the cyclic construction method for $(h-1)$-spreads of $\mathcal{P}_u$ is widely-used, it accesses only a fraction of all possible spread/star-based designs. For example, only one of the 131044 isomorphism classes of $1$-spreads of $\mathcal{P}_6$ is obtained using the cyclic method. As per our knowledge, one may have to rely on exhaustive search to find non-isomorphic spread/star-based designs.


\section{Concluding Remarks}

In this paper, we formalize the definition of isomorphism of multi-stage factorial designs under the unified framework based on randomization defining contrast subspace (RDCSS), developed by \cite{ranjan2007factorial}. Focussing on the RDCSS-based designs that are derived from balanced spreads and balanced covering stars, we have developed  isomorphism check algorithms that are more efficient than the naive approach of iterating through all possible relabelings and reorderings. We have also provided a complete classification of small designs that are typically assumed to be important from practical standpoint. Furthermore, the proposed algorithms and relevant functions are implemented in \verb"R" for easy access.

A few remarks are as follows. Both the theoretical results and the algorithms can easily be generalized for unbalanced spreads and stars, however, the construction and complete classification of such designs require more work. Some of the theoretical results will also hold when generalized to $q$-level multi-stage factorial designs with randomization restrictions. The proposed relabeling approach can also be used to find a design that meet the pre-specified randomization restrictions.

\begin{acknowledgements}
The authors would like to thank the chief editor, the handling editor, and the reviewers for their helpful comments. 
\end{acknowledgements}

%
 \section*{Conflict of interest}
 The authors declare that they have no conflict of interest.



\bibliographystyle{spbasic} 
\bibliography{ref2018}

\newpage
\section*{Appendix~A: Proofs of Results in Section~4}

\noindent{\emph{Proof of Proposition~\ref{propsize}}.} Without loss of generality, suppose that $f_{1}, \ldots, f_{\ell}$ are the RDCSSs for $d_1$ that contain at least one of $x_1,\ldots, x_n$. 
Then, by the partial RDCSS mapping property, the elements of distinct $f_{1}, \ldots, f_{\ell}$ must be mapped to distinct RDCSSs in $d_2$. 
There are $\mu!/(\mu-\ell)!$ different ways to choose a correspondences between these $\ell$ RDCSSs in $d_1$ and $\ell$ of the $\mu$ RDCSSs which comprise $d_2$. Subsequently, there are $\prod_{j=1}^{m_i} \left( 2^t-2^{j-1} \right)$ distinct choices of linearly independent points in each RDCSS of $d_2$ to which we can map the $m_i$ points $f_i$. Combining these counts as a product yields the result.\\

\noindent{\emph{Comment on Proposition~\ref{propsize}}.}  The upper bound given in Proposition~\ref{propsize} is not necessarily tight--- it is possible that some RDCSS correspondences do not yield full rank solutions. For example, if $n =6$, $m_1 = m_2 = m_3 = 2$ and $g_1 = \langle A, B\rangle$, $g_2 = \langle C, D\rangle$, $g_3 = \langle AC, BD\rangle$, then no collineations exist for the RDCSS correspondence $(f_1, f_2, f_3) \rightarrow (g_1, g_2, g_3)$ because $\langle g_1\cup g_2 \cup g_3 \rangle$ is not full rank. Therefore, this entire correspondence can be discarded from the search, providing an even greater reduction. \\

\noindent{\emph{Proof of Proposition~\ref{itspossible}}.} When $t=1$, the result is trivial, so we limit our consideration to $t > 1$. Suppose that within $\psi$ there exists $k$ $(t-1)$-flats $f_{u_1},\ldots, f_{u_k}$ of $\mathcal{P}_n$ such that $|\langle \cup_{i=1}^{k} f_{u_i} \rangle| = 2^{kt}-1$ for some integer $k \leq n/t$. This is guaranteed to at least hold for $k=1$ by definition of a spread. If $k =n/t$, then the result is immediate. Otherwise, $k \leq n/t - 1$, and of the $2^n - k 2^t$ points not contained within $\cup_{i=1}^{k} f_{u_i}$, $2^n - 2^k$ are not contained by $\langle \cup_{i=1}^{k} f_{u_i} \rangle $ leaving $2^k - k 2^t$ that do fall within that span. Recall that all $(t-1)$-spreads of $\mathcal{P}_n$ contain $\mu = (2^n-1)/(2^t-1)$ $(t-1)$-flats. Then, $|\psi| - k$ is given by
\begin{eqnarray*}
	\mu - k = \frac{2^n-1}{2^t-1} -k = \sum_{i=1}^{n/t} 2^{(i-1)t} - k \geq& 2^{k} - k.
\end{eqnarray*}
The pigeonhole principle guarantees that at least one $(t-1)$-flat $f_{u_{k+1}}$ shares no elements with $\langle \cup_{i=1}^{k} f_{u_i} \rangle $. 
This flat can be appended to the list $f_{u_i},\ldots, f_{u_k}$ without introducing any linear dependence. Proceeding inductively, additional flats can be included until $k = n/t$. $\Box$\\


\newpage
\section*{Appendix~B: Proofs of Results in Section~6.2}

First we give a technical lemma and then the proofs of the two theorems.

\begin{lemma}\label{lemma:123}
Let $\psi=\{f_1,...,f_{\mu}\}$ be an $(h-1)$-spread of $\mathcal{P}_u$ constructed with the cyclic method using a primitive polynomial $P(\omega)$ and root $\omega$. Then,
\begin{enumerate}[(a)]
     \item $x_1 = \omega^a$ and $x_2 = \omega^b$ are in the same $(t-1)$-flat $f\in\psi$ if and only if $a \equiv b  \bmod \mu$;
     \item the set of all nonzero roots of $\omega^{2^h} - \omega$ is equal to the set of all elements of the form $\omega^a$ where $a \equiv 0 \bmod \mu$.  Thus the first $(h-1)$-flat, $f_1\in\psi$, corresponds to the set of all nonzero elements of $GF(2^h)$;
     \item $f_2,...,f_{\mu}$ are multiplicative cosets of $f_1$ in the group $GF(2^u)^*$ of nonzero elements of $GF(2^u)$.
\end{enumerate}
\end{lemma}

\noindent{\emph{Proof}.} (a) follows trivially from the cyclic structure in Table~\ref{tab:cycle}. For part (b), since $\mu (2^h - 1) = 2^u - 1$, or $\mu 2^h \equiv \mu \bmod 2^u - 1$,
$$
    (\omega^{\ell \mu})^{2^h} = ( \omega^{\mu 2^h})^\ell = (\omega^\mu)^\ell = \omega^{\ell \mu},
$$
and hence, $\omega^{\ell \mu}$ is a root of $\omega^{2^h} - \omega$. Part (c) follows from noting that the elements of $f_{i}$ are of the form $\omega^{k \mu + i} = \omega^i \omega^{k \mu}$, where $0 \le i < \mu$. $\Box$

\bigskip

\noindent{\emph{Proof of Theorem \ref{thm:cyclicequiv}}.} We need to show that for every $g_j \in \psi_2$, there exists a unique $f_i \in \psi_1$ such that the elements in $g_j$ are in $f_i$. Let $e_1$ and $e_2$ be two distinct effects in $g_j$, then from Lemma~\ref{lemma:123}(a), $e_{1} =\beta^a$, $e_{2} = \beta^b$ and $a \equiv b$ (mod $\mu$). From  Theorem~2.14 of \cite{lidl1994introduction}, there exists $0 \le k\le u$ such that $\beta=\alpha^{2^k}$. Thus, $e_{1} = \beta^a = (\alpha^{2^k})^a=\alpha^{2^ka}$ and $e_{2} = \beta^b = \alpha^{2^kb}$. Note that $a \equiv b$ (mod $\mu$) implies $2^ka \equiv 2^kb$ (mod $\mu$), as $gcd(2^k,\mu) = 1$. Consequently, $e_1$ and $e_2$ must belong to the same flat in $\psi_1$ (from Lemma~\ref{lemma:123}(a)). $\Box$ \\

\bigskip

\noindent{\emph{Proof of Theorem \ref{thm:cyclic_iso}}.}  We establish the existence of an IEC by constructing one.  Our isomorphism will be a field isomorphism, which makes it easier to show that it is an IEC.

Let $\alpha$ be the primitive root of $P_1(\omega)$ which is used to construct $\psi_1$ and let $\beta$ be the primitive root of $P_2(\omega)$ which is used to construct $\psi_2$.
By \citet[Thm 2.40]{lidl1994introduction}, there is a primitive polynomial $Q(x)$ of degree $u$ whose roots form a basis for $GF(2^u)$ over $\mathbb{Z}_2$.
Note that if $\omega$ is one of these roots then the other $u-1$ roots are all of the form $\omega^{2^i}$ for $i=1,\ldots, u-1$.
There are $a,b \in \{1,\ldots, 2^u-2 \}$ with both $\alpha^a$ and $\beta^b$ roots of $Q(x)$. We define our IEC $\Phi$ by first setting
\[
      \Phi( (\alpha^a)^{2^i}) = (\beta^b)^{2^i} \quad \mbox{ for } i = 0, 1, \ldots, u-1,
\]
and then extending $\Phi$ to all of $GF(2^u)$ by linearity. Since the roots of $Q(x)$ form a basis, this uniquely defines $\Phi$.

Our next task is to show that $\Phi$ is a field isomorphism. By our definition, $\Phi$ is linear; we need only show that $\Phi$ preserves multiplication.
Since $Q(x)$ is primitive and $\alpha^a,\beta^b$ are both roots of $Q(x)$,  it is enough to show $\Phi( (\alpha^a)^k) = (\beta^b)^k$ for all $k=1,\ldots, 2^u-1$. Fix $k$.  Since $\alpha^a, \alpha^{2 a}, \ldots, \alpha^{2^{u-1} a}$ are the distinct roots of $Q(x)$ and are a basis, there are constants $c_i \in \mathbb{Z}_2$ so that
\[
     \alpha^{a k} = \sum_i c_i \alpha^{a 2^i}.
\]
Consider the polynomial $H(x) = x^k - \sum_i c_i x^{2^i}$. Then $H(\alpha^a) = 0$ by definition of $c_i$. However, since $x \mapsto x^{2^j}$ is a field automorphism for any $j$, this means that $H(\alpha^{a 2^j}) = 0$ as well. Thus all the roots of $Q(x)$ are also roots of $H(x)$. Since $\beta^b$ is a root of $Q(x)$, then $H(\beta^b) = 0$ or $\beta^{b k} = \sum_i c_i \beta^{b 2^i}$.  However, then
\begin{eqnarray*}
    \Phi(\alpha^{ak}) &=& \Phi( \sum_i c_i \alpha^{a 2^i} ) = \sum_i c_i \Phi( \alpha^{a 2^i}) \cr
                                 &=& \sum_i c_i (\beta^b)^{2^i} = \beta^{b k}
\end{eqnarray*}
and so $\Phi$ is also a field isomorphism.   We claim that $\Phi$  is an IEC.  To see this, we first note that by \citet[Thm 2.21]{lidl1994introduction}, $\Phi$ maps the roots of
$x^{2^h} - x$ to roots of $x^{2^h} - x$.
That is, it maps $GF(2^h) \subset GF(2^u)$ to itself.
This indicates, by Lemma \ref{lemma:123}(b),  that $\Phi(f_1) = g_1$.
Additionally, since $\Phi$ is a field isomorphism, it maps any multiplicative coset of $GF(2^h)^*$ in $GF(2^u)^*$ to some multiplicative coset of $GF(2^h)^*$.
By Lemma \ref{lemma:123}(c), each $(h-1)$-flat $f_i$ of $\psi$ is mapped to a unique $(h-1)$-flat $g_j$ in $\psi_2$.
Thus $\Phi$ is an IEC for $\psi_1$ and $\psi_2$.
$\Box$ \\


\newpage
\section*{Appendix~C: R Codes for Easy Implementation}

In this section, we discuss various functions that are used to implement Algorithm~1 and Algorithm~2 for checking the isomorphism of balanced spread- and star- based designs. These functions are implemented in R and have been uploaded to GitHub for easy access (see https://github.com/neilspencer/IsoCheck/). The usage and brief description of the key functions are as follows:

The isomorphism of two $(t-1)$-spreads of $PG(n-1, 2)$, \verb"spread1" and \verb"spread2", can be checked using the following R code:
\begin{verbatim}
     R> checkSpreadIsomorphism(spread1, spread2, returnfirstIEC = T).
\end{verbatim}
The third argument \verb'"returnfirstIEC = T"' specifies whether the algorithm searches until it finds the first IEC (might only take a few second) or if it continues to search for and returns all IECs (which can take a long time). For non-isomorphic spreads or stars, the run times are the same (none are found). However, for isomorphic spreads, stopping once we have found one IEC (which means they are isomorphic) is much faster.

Similar to spread-isomorphism, two stars \verb"star1" and \verb"star2" can be checked for isomorphism using the following R code

\begin{verbatim}
     R> checkStarIsomorphism(star1, star2, returnfirstIEC = T).
\end{verbatim}

It is assumed that both spreads are $(t-1)$-spreads, and both stars are $St(n, \mu, t, t_0)$ of $PG(n-1, 2)$. The isomorphism check for stars is slightly different than for spread --- it exploits the spread to star correspondence to reduce the dimension of the search space (as described by Algorithm~2). Both \verb"checkSpreadIsomorphism" and \verb"checkStarIsomorphism" call several important functions such as finding the bitstring representation of flats for checking equivalence, and applying collineations for relabeling of spreads and stars. The usage of these functions are illustated as follows:

\begin{verbatim}
     R> getBitstrings(spread1)
     R> applyCollineation(C, spread1)
     R> checkspreadEquivalence(spread1, spread2) 
     R> checkstarEquivalence(star1, star2) 
\end{verbatim}

Though the user can input the spreads of their choice in a specified format as discussed in the \verb'"readme"' file and \verb'"exampleScript.R"', we have coded several spreads and stars that are used in this paper (see the help manual of the R package \verb'"IsoCheck"').

\end{document}